\documentclass{article}

	\usepackage{amsmath}
	\usepackage{epsfig}
	\usepackage{color}
	\usepackage{float}
	\usepackage[table]{xcolor}
	\usepackage[margin=1.5in]{geometry}
	\usepackage[utf8]{inputenc}
	\usepackage{siunitx}
	\usepackage{textgreek}
	\usepackage{upgreek}
	\usepackage{graphicx}
	\usepackage{subcaption}
	\usepackage[toc,page]{appendix}
	\usepackage[affil-it]{authblk}
	\usepackage[english]{babel}
	\usepackage{blindtext}
	\usepackage{cite}

	\newcommand{\bd}{~\mathbf{d}}
	\newcommand{\bRp}{\mathbf{R}^+}
	\newcommand{\bRn}{\mathbf{R}^-}
	\newcommand{\bR}{\mathbf{R}}
	\newcommand{\brp}{\mathbf{r}^+}
	\newcommand{\brn}{\mathbf{r}^-}  
	\newcommand{\br}{\mathbf{r}}
	\newcommand{\be}{~\mathbf{e}}
	\newcommand{\pardev}[2]{\frac{\partial{#1}}{\partial{#2}}}
	\newcommand{\bP}{\mathbf{P}}
	\newcommand{\bQ}{\mathbf{Q}}
	\newcommand{\bZ}{\mathbf{Z}}

	\newcommand{\bin}{\textbf{\textit{n}}}

	\newcommand{\bim}{\textbf{\textit{m}}}

	\newcommand{\bw}{\mathbf{w}}

	\newcommand{\bc}{\textbf{\textit{c}}}
	\newcommand{\bfo}{\textbf{\textit{f}}}
	
	\newcommand{\bfv}{~\mathbf{f}}
	
	\newcommand{\rhop}{\omega x}

	\newcommand{\cosr}{\cos \omega x}
	\newcommand{\sinr}{\sin \omega x}

	\newcommand{\er}{(\cosr\be_1+\sinr\be_2)}

	\newcommand{\tp}{\mathbf{t}^+}
	\newcommand{\np}{\mathbf{n}^+}
	\newcommand{\bp}{\mathbf{b}^+}
	
	\newcommand{\tpz}{\mathbf{t}^+_0}
	
	\newcommand{\bpz}{\mathbf{b}^+_0}

	\newcommand{\tn}{\mathbf{t}^-}
	\newcommand{\nn}{\mathbf{n}^-}
	\newcommand{\bn}{\mathbf{b}^-}

	\newcommand{\oeps}{O(\varepsilon)}
	\newcommand{\oepss}{O(\varepsilon^2)}
	
	\newcommand{\thetap}{{\Theta^+}}
	\newcommand{\thetan}{{\Theta^-}}

	\newcommand{\phic}{\pmb{\phi^c}}
	
	\newcommand{\boeta}{\pmb{\eta}}

	\newcommand{\negs}{^-}
	\newcommand{\poss}{^+}
	\newcommand{\spm}{^\pm}

	\newcommand{\bv}{\mathbf{v}}

	\newcommand{\eqsp}[1]{\begin{equation}
			\begin{split}
				#1
			\end{split}
	\end{equation}}

	\newcommand{\bu}{\mathbf{u}}

	\newcommand{\by}{\mathbf{y}}

	\newcommand{\upc}{^c}

		\title{Elasticity as the basis of allostery in DNA}
		\date{}
		\author[1]{Jaspreet Singh}
		\author[1,*]{Prashant K. Purohit}
		
		\affil[1]{Department of Mechanical Engineering \& Applied Mechanics,
			University of Pennsylvania, Philadelphia, PA 19104.}
		\affil[*]{Corresponding author: Prashant K. Purohit, purohit@seas.upenn.edu}
		
\begin{document}
			\maketitle
		\begin{abstract}
			Allosteric interactions in DNA are crucial for various biological processes. These interactions are quantified by measuring the change in free energy as a function of the 
			distance between the binding sites for two ligands. Here we show that trends in the interaction energy of ligands binding to DNA can be explained within an elastic birod model. The birod model accounts for
			the deformation of each strand as well as the change in stacking energy due to perturbations in position and orientation of the bases caused by the binding of ligands. The strain fields
			produced by the ligands decay with distance from the binding site. The interaction energy of two ligands decays exponentially with the distance between them and oscillates with the periodicity of the double helix in quantitative agreement with experimental measurements. The trend in the computed interaction energy is similar to that in the perturbation of groove width produced by the binding of a single ligand which is
			consistent with molecular simulations. Our analysis provides a new framework to understand allosteric interactions in DNA and can be extended to other rod-like macromolecules whose elasticity plays a role 
			in biological functions.  
		\end{abstract}
		

		
\begin{figure}[!]
	\centering{
		\includegraphics[width=4.5cm,height=3.5cm]{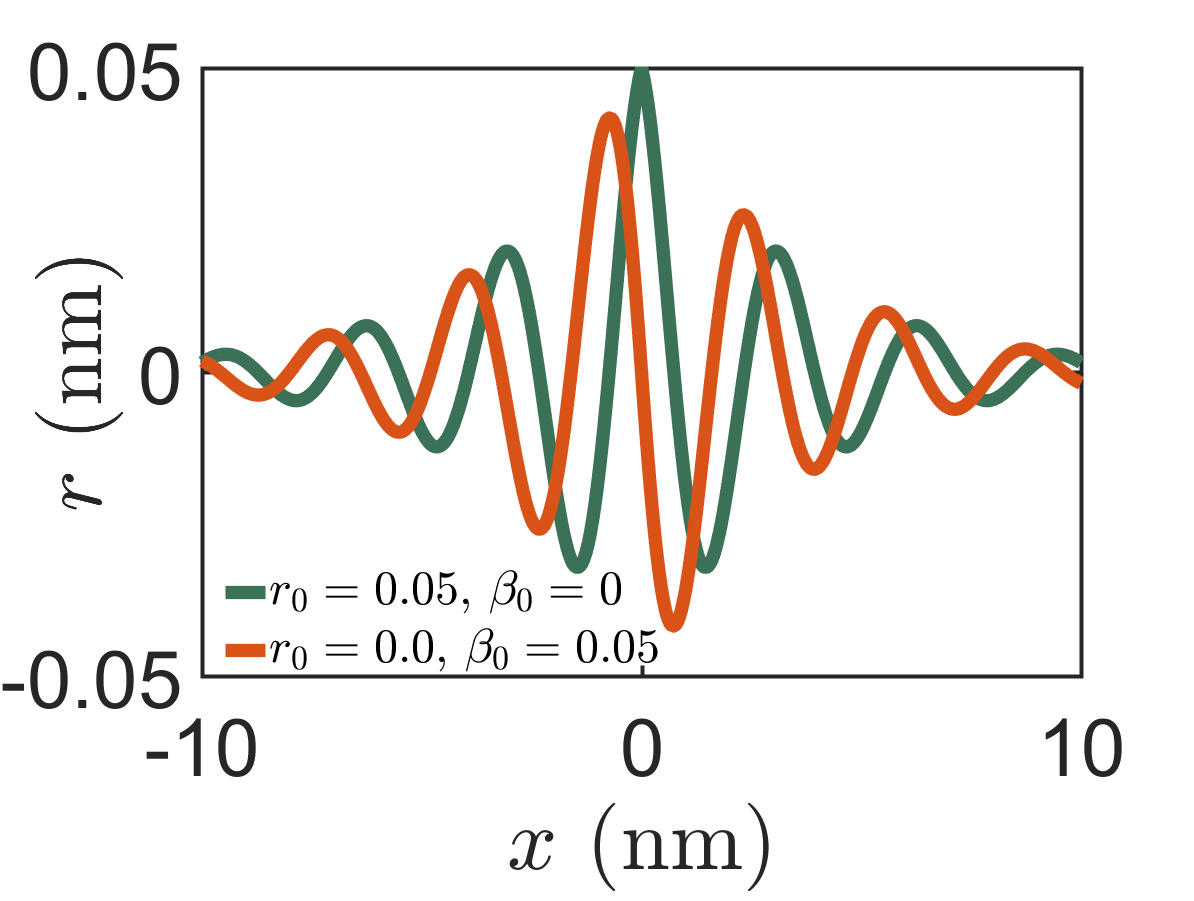}
		\includegraphics[width=4.5cm,height=3.5cm]{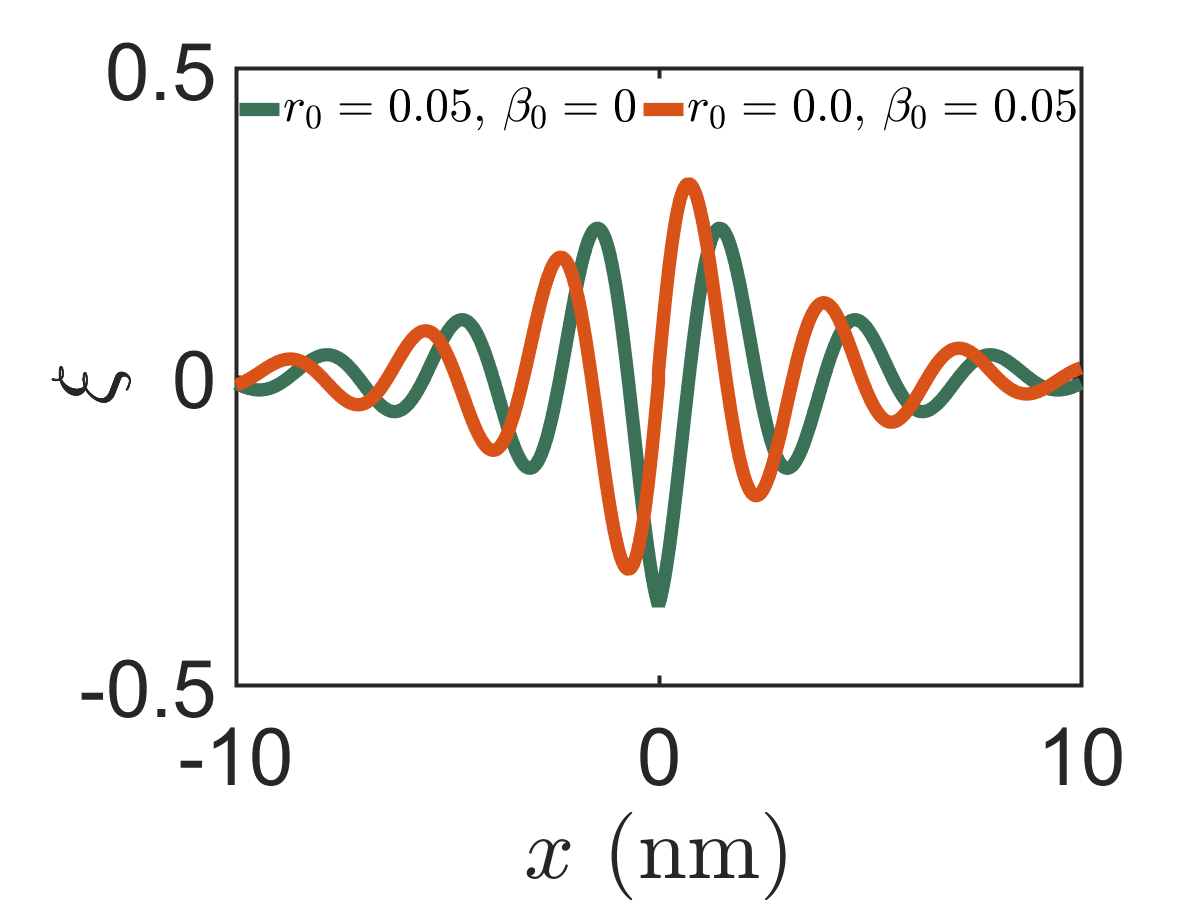}
		\includegraphics[width=4.5cm,height=3.5cm]{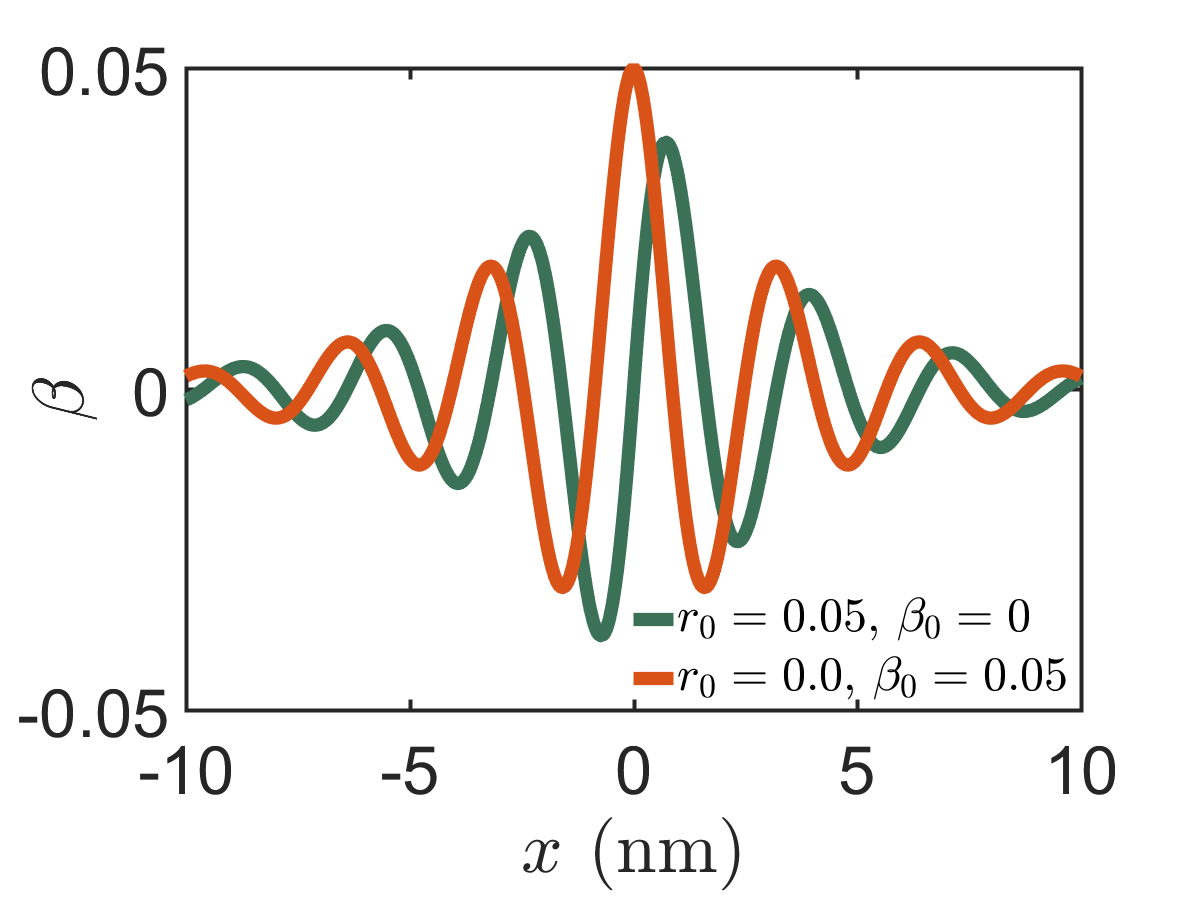}
		\caption{Variation of $r,k_3,\xi$ and $\beta\poss=\beta\negs=\beta$ for a single protein. The red curve corresponds to the boundary conditions $\beta_0=0,r_0=0.05$ nm and the green curve to $r_0=0,\beta_0=0.05$. The decay length is $l_d=\zeta^{-1}\approx10$ bp which is close to that documented in literature \cite{kim_science,xu2013modeling}. }
		\label{fig:1prot1}}
\end{figure}

\begin{figure}[!]
	\centering{
		\includegraphics[width=8cm,height=6cm]{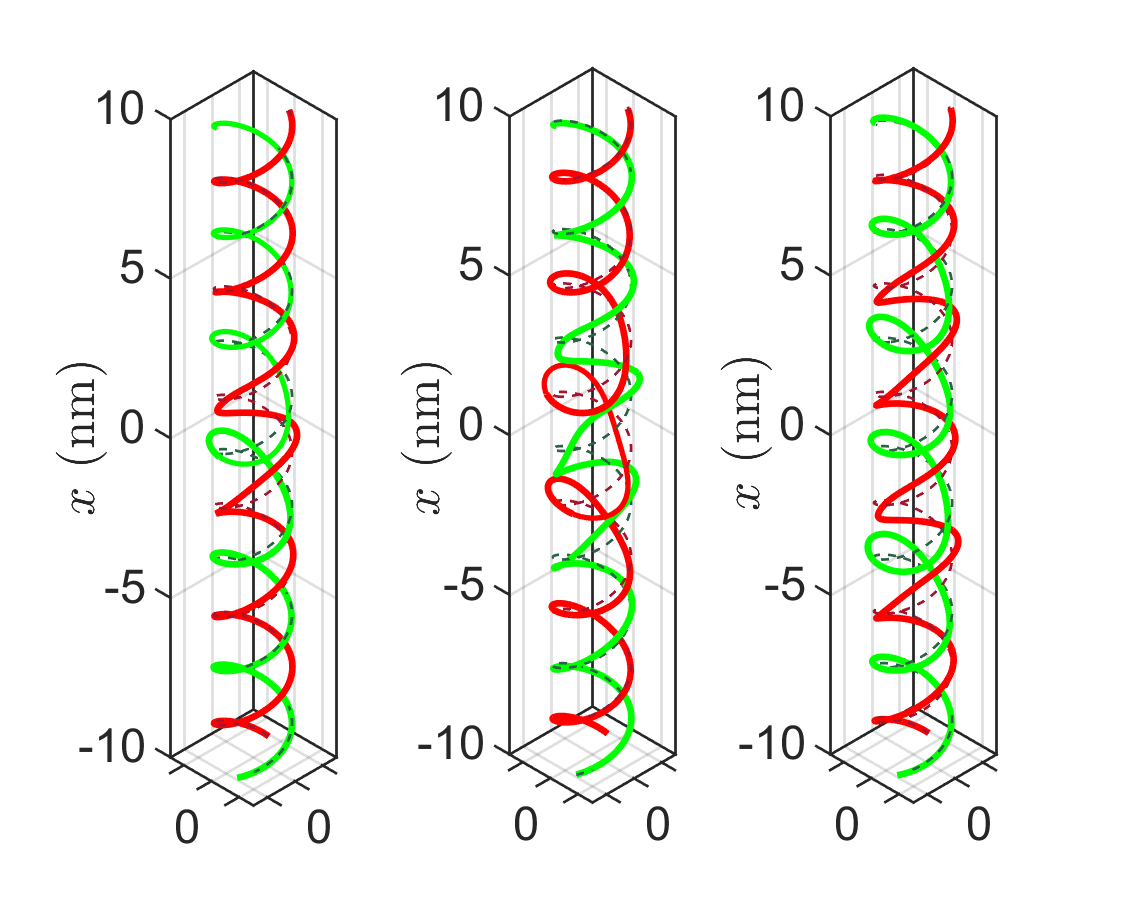}
		\caption{We show the deformed configuration of the double helix, red and green colors correspond to $+$ and $-$ strand, respectively. In the first figure, one protein binds at $x=0$ with $r_0=0.2$ nm and $\beta_0=0$. In the second figure, two proteins bind at $x=\pm 1.5$ nm. In the third figure, two proteins bind at $x=\pm 3.5$ nm. Notice the overlap of deformations in the second figure which is absent in the third one. This overlap is manifests itself as interaction energy between the two proteins. The dotted lines denote the corresponding undeformed configuration. }
		\label{fig:helixshape}}
\end{figure}

\begin{figure}[!]
	\centering{
		\includegraphics[width=4.5cm,height=3.5cm]{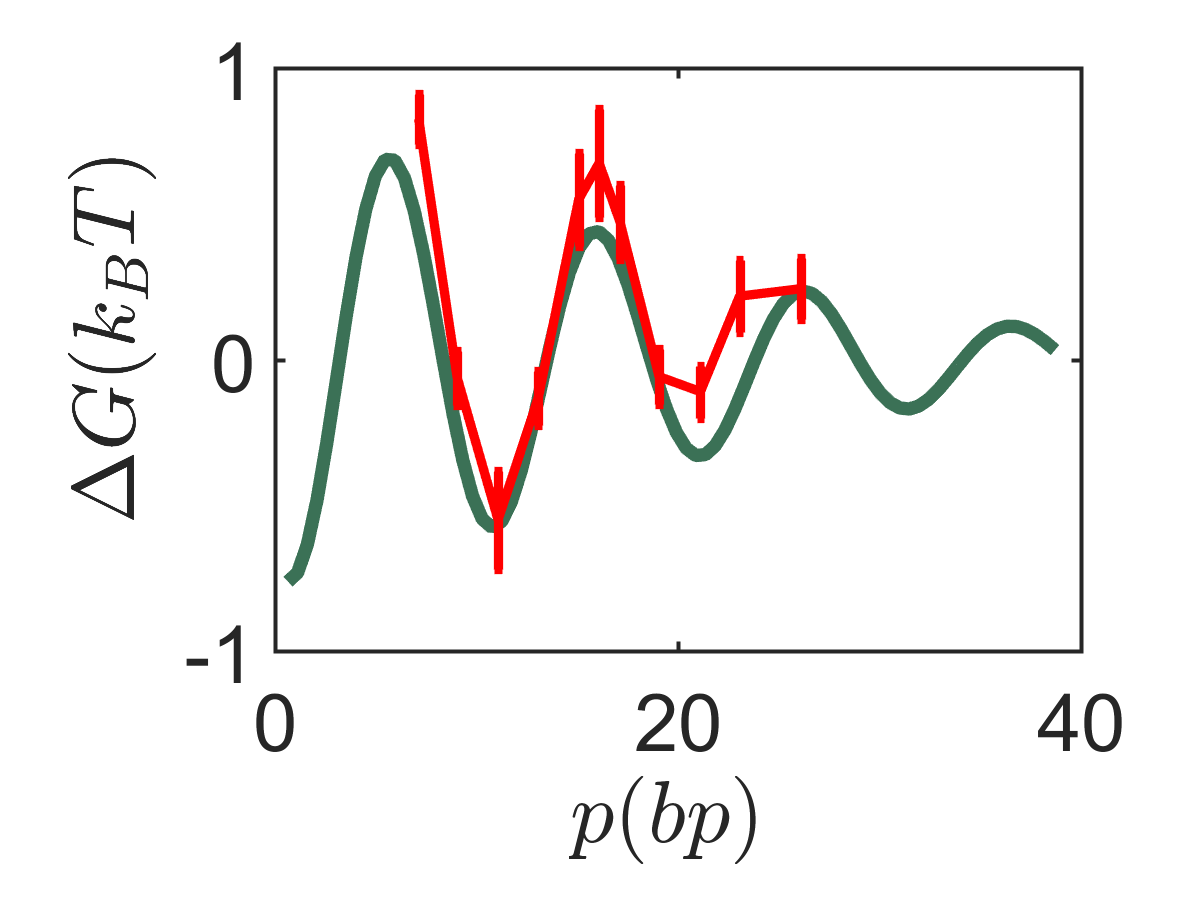}
		\includegraphics[width=4.5cm,height=3.5cm]{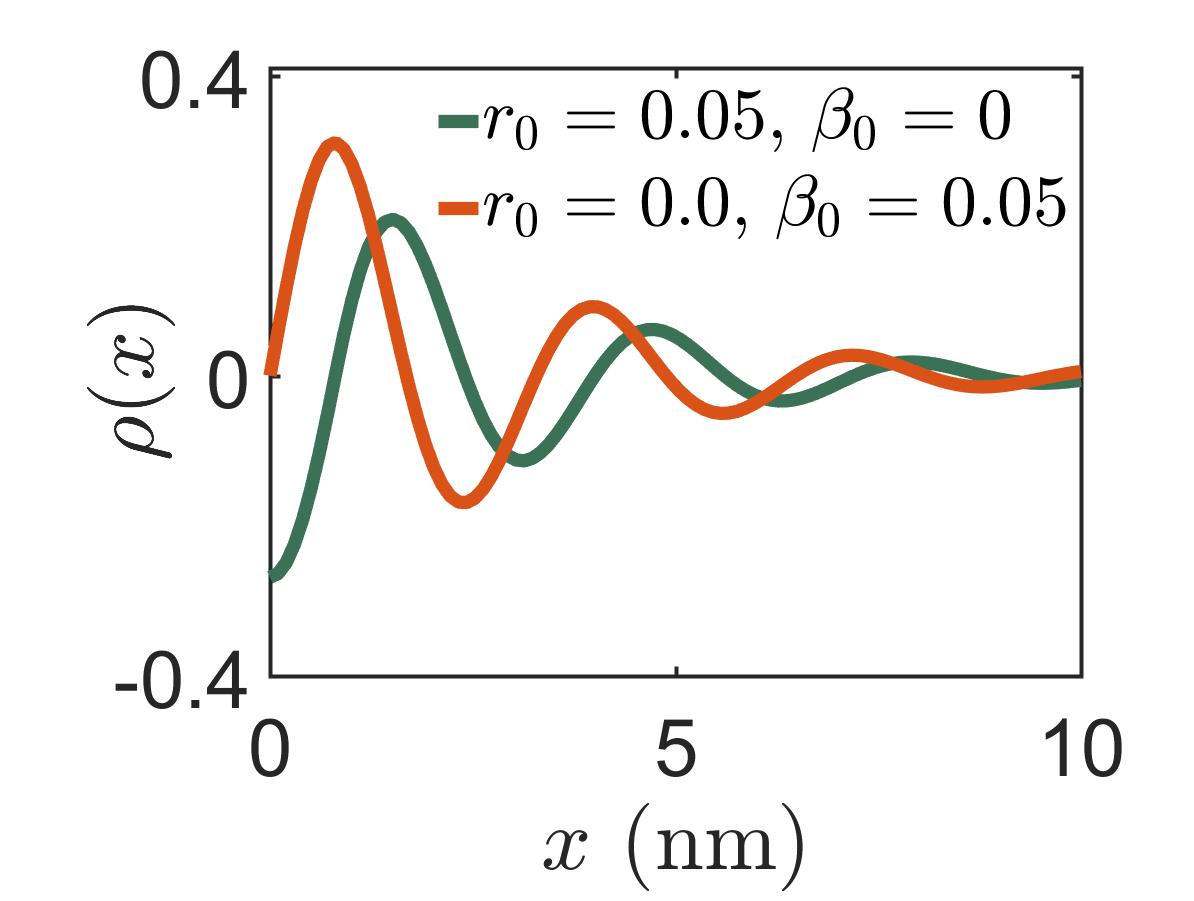}
		\caption{The first figure shows the variation of interaction energy $\Delta G$ with distance $p$ between the two proteins $P_1$ and $P_2$. The boundary conditions $r_1=0.001$ nm, $\beta_1=0.0045$ for $P_1$ and $r_2=0.001$ nm, $\beta_2=-0.0045$ for $P_2$ give the best fit to the experimental data for $\Delta G$\cite{kim_science}. In the second figure, we show the variation of change in groove width $\rho(x)=g(x)-\frac{p}{2}$ when a protein with boundary conditions $r_0,\beta_0$ binds at $x=0$. The decaying sinusoidal character is documented in previous work \cite{kim_science}\cite{gu2015dna}. The magnitude of the change in groove width ($\sim 3$ A) is consistent with estimates in \cite{kopka1985molecular}.}
		\label{fig:DeltaGandrho}\vspace{-0.5cm}}
\end{figure}

\vspace*{1cm}
When a ligand binds to DNA it induces conformational changes at the binding site which could propagate to regions tens of base-pairs away, thereby encouraging or inhibiting the binding of a second ligand in those places. Such interactions between two binding agents are called {\it allosteric} interactions. Our focus here is on a mechanism for allostery based on elasticity of long molecules. Although we will illustrate our theory using DNA as an example, long range allosteric interactions have been documented in actin, microtubules and helical peptide chains. For example, myosin binds to actin filaments leading to suppression of the formation of cofilin clusters via allosteric signalling \cite{ngo2016allosteric}. Long range structural changes induced by taxol binding to microtubules inside a cell prevents cell division thus making it a potent anti-tumor agent \cite{mitra2008taxol}. The transfer of chiral stimulus triggered by a binding agent across a helical peptide chain gives the molecule an overall chiral character and is yet another instance of allostery \cite{ousaka2008transfer}. We will analyze allostery in dsDNA because detailed experimental and simulation results are available for it~\cite{drsata2014mechanical,kim_science,koslover2009twist}, thus allowing quantitative comparisons with our theory. \\  
DNA comprises of two helical strands held together via complementary base-pairing. When a ligand, such as a protein or a drug, binds to DNA it exerts forces and moments on the double helix \cite{efremov2018transfer,wiggins2009protein} causing deformations at the base-pair level. We use the theory of birods \cite{maddocks_birod} to investigate these deformations. A birod consists of two elastic strands which interact through an elastic web. This construction makes it suitable for investigating the deformations at the base-pair level in a DNA molecule which a homogeneous rod model cannot capture~\cite{lankavs2009parameterization}. The latter ignores the double helical structure and the elasticity of the base pairs, both of which are crucial to the problem under consideration.
\begin{figure}[]
	\centering{
		\includegraphics[width=6cm,height=5cm]{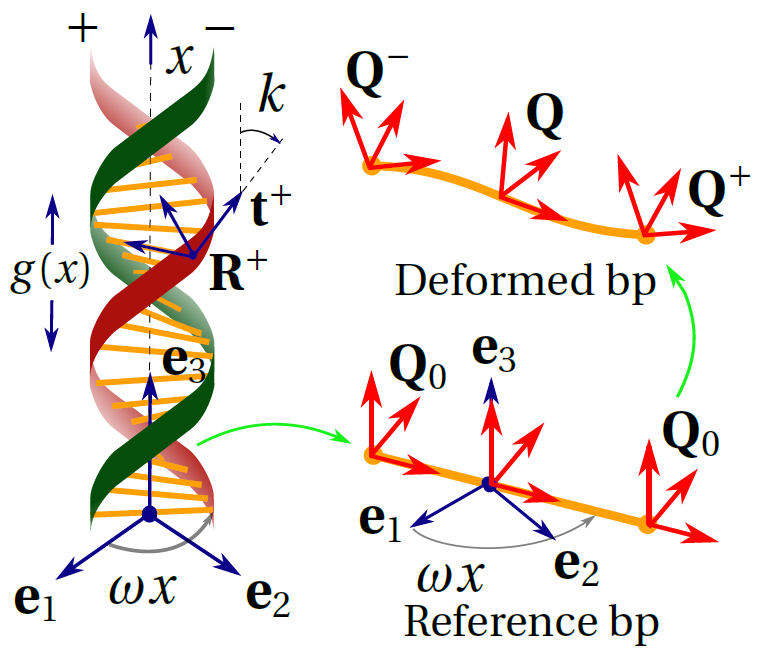}
		\caption{Birod model of DNA. The angle between the tangent $\textbf{t}\poss$ and $\be_3$ is $k$. A base pair in reference and deformed state is shown. The director frames attached to $\pm$ ends of the base pair change from $\bQ_0$ to $\bQ\spm$, respectively. The rigid rotation of the strand $\bQ=(\bQ\poss\bQ^{-T})^{\frac{1}{2}}\bQ\negs$ and micro-rotation $\bP=(\bQ\poss\bQ^{-T})^{\frac{1}{2}}$.}
		\label{fig_DNA}}
\end{figure}
In this letter, $(\cdot)_x$ denotes $\pardev{(\cdot)}{x}$. Lower case letters such as $a,r,\beta^\pm$ are scalars, bold lower case letters such as $\tp,\bn$ are vectors while bold upper case letters such as $\bRn,\bRp_0,\bZ$ are $3\times 3$ tensors.\\ We assume the phosphate backbones comprising of phosphodiester bonds to be inextensible and unshearable elastic strands. Since these backbones consist of consecutive single bonds which allow for free rotation about the bond, we assume that they can not resist twisting moments. The base pairing is represented by the elastic web which is capable of extending, shearing, bending and twisting. In addition to the elastic energy, we consider contributions from the stacking energy which is associated with the change in orientations of the successive base pairs.\\
We denote the helical strands as $\pm$; their positions in the reference state are denoted by $\br^\pm_0$. We use arclength parameter $x$ to parametrize the double helix. Thus, 
\eqsp{ \brp_0&=a(\cos \omega x\be_1+\sin \omega x\be_2)+x\be_3,\\
	\brn_0&=a(\cos (\omega x+\alpha)\be_1+\sin (\omega x+\alpha)\be_2)+x\be_3,
	\label{Eqn:ref_helix}
}where $a=1$ nm is the radius of the DNA helix, $p=3.4$ nm is the pitch, $\omega=\frac{2\pi}{p}$ and $\alpha$ is the phase difference between the helices. Here we assume $\alpha=\pi$ to make the computations analytically tractable. We consider a deformed configuration where the double helix extends and twists about $\be_3$, and its radius and phase angle also change due to binding of ligands. The deformed state of the $\pm$ strands is denoted by $\br^\pm(x)$, where
\eqsp{
	&\brp(x)=(a+r)\Big(\cos(\rhop+\beta\poss)\bd_1+\sin(\rhop+\beta\poss)\bd_2\Big)+(x+\int_{-\infty}^{x}a\xi\poss dx)\be_3,\\
	&\brn(x)=-(a+r)\Big(\cos(\rhop+\beta\negs)\bd_1+\sin(\rhop+\beta\negs)\bd_2\Big)+(x+\int_{-\infty}^{x}a\xi\negs dx)\be_3,
	\label{Eqn:def_helix}}
%
%
such that $\bd_{1x}=k_3 \bd_2$ and $\bd_{2x}=-k_3 \bd_1$. We assume all the displacement and strain parameters $r,\beta^\pm$ and $\xi^\pm$ vanish at $x=\pm\infty$ because the deformations caused by the proteins are local. The change in radius $r$, change in the phase angle $\beta^\pm$, stretch $\xi$, and the twist $k_3$ are assumed to be small ($\sim \oeps$) such that second order terms such as $r^2$ and $\xi\beta\negs $ are negligible. However, there could be finite rotations resulting from $k_3$.\\  
We need to solve the balance equations for a birod \cite{maddocks_birod}, which are 
\eqsp{
	&\bin^\pm_x\mp\bfo+\textbf{\textit{l}}=0,\\
	&\bim^\pm_x+\br^\pm_x\times\bin^\pm+\frac{1}{2}(\brp-\brn)\times\bfo\mp\bc+\textbf{\textit{h}}=0,
	\label{Eq:gov_eqn}
}
where $\textbf{\textit{m}}\spm$ and $\textbf{\textit{n}}\spm$ are the contact moment and contact force respectively in $\pm$ strands. $\bfo$ and $\bc$ are the distributed force and distributed moment exerted by the $+$ strand on the $-$ strand. $\textbf{\textit{l}}$ and $\textbf{\textit{h}}$ are the body force and body moment exerted by the base pairs onto both $\pm$ strands. We use the position vectors for the deformed helix $\br\spm(x)$ (eqn. \ref{Eqn:def_helix})  to compute these quantities.\\    
The outer strands are inextensible which means $|\br_x^\pm|=|\br_{0x}^\pm|$ yielding,
\eqsp{
	\omega^2 r +a\omega (k_3+\beta^\pm_x)+\xi^\pm=0.
	\label{Eqn:inext}
} We use the above equation to eliminate $\xi\spm$. We attach a director frame $\bR^\pm=[\textbf{n}^\pm_0\quad \textbf{b}^\pm_0\quad\textbf{t}^\pm_0]$ to each cross-section of the $\pm$ strands, where $\textbf{n}^\pm_0$, $\textbf{b}^\pm_0$, and $\textbf{t}^\pm_0$ are the normal, binormal, and tangent in the reference state to $\pm$ strand, respectively. $\textbf{n}^\pm,\textbf{b}^\pm_0,\textbf{t}^\pm_0$ and the curvature in the reference configuration $\Omega^\pm_0(=\omega \sin k)$ are computed using eqn. $\ref{Eqn:ref_helix}$ (see the supplement). Similarly, we use eqn.\ref{Eqn:def_helix} to compute the the Frenet-Serret frame $\bR^\pm=[\textbf{n}^\pm\quad \textbf{b}^\pm\quad\textbf{t}^\pm]$ and curvature $\Omega^\pm$ in the deformed state. We neglect terms higher than first order such as $r\beta\poss,\xi\negs r\sim \oepss$ and summarize the results in eqn. \ref{Eq:Rframes}. The bending moment in the outer strands $\bim\spm$ is proportional to the change in curvature $\kappa\spm=\Omega\spm-\Omega_0^\pm$ and is directed along the binormal $\textbf{b}\spm$ such that $\bim\spm=EI\kappa\spm\textbf{b}\spm$ where $EI$ is the bending modulus of the strand. Note that the twisting moment is zero.\\
\eqsp{
	& \Omega \spm=\Omega\spm_0-(r_{xx}+\xi\spm)\cos k+(\beta\spm_{x}+k_3)\sin k,\\
	& \bR\spm=[\textbf{n}\spm\quad\textbf{b}\spm\quad\textbf{t}\spm]=\mathbf{Z}\bR\spm_0(\mathbf{1}+\Theta\spm),\\
	& \bZ=\bd_1\otimes\be_1+\bd_2\otimes\be_2+\be_3\otimes\be_3,\\
	&\Theta\spm=\begin{bmatrix}
		0 &-\theta\spm_3&\theta\spm_2\\\theta\spm_3&0&-\theta\spm_1\\-\theta\spm_2&\theta\spm_1&0
	\end{bmatrix}
	\label{Eq:Rframes}}
\begin{align*}
	&\theta\spm_1=r\omega+a(\beta\spm_x+k_3),\quad \theta\spm_2=-r_x\cos k+\beta\spm \sin k,\\
	&\theta\spm_3=\frac{-\omega r_x -a(\beta\spm_{xx}+k_{3x})}{\omega \sin k}-\frac{(r_x\cos k-\beta\spm \sin k)\cos k}{\sin k}.
\end{align*}  
Now, we compute the bending and twisting of the web which represents base-pairing. We attach a director frame $\bQ_0$ to both $+$ and $-$ end of the base pair (fig.\ref{fig_DNA}). 
\eqsp{\bQ_0=[\be_r\quad\be_\theta\quad\be_3],}
where $\be_r=\cosr\be_1+\sinr\be_2$ and $\be_\theta=-\sinr\be_1+\cosr\be_2$. As the birod deforms, these frames respectively get mapped to $\bQ\spm$. We compute $\bQ\spm$ using the deformation of $\bR\spm$ from eqn. (\ref{Eq:Rframes}).
\eqsp{
	\bQ\spm=\bZ\bR_0\spm(\textbf{I}+\Theta\spm)\bR_0^{\pm T} \bQ_0,\quad\quad \Theta\spm\sim \oeps.
} 
Now, we can compute the rigid rotation $\bQ$ and micro-rotation $\bP$ for each base pair. The micro-rotation contains information about the `difference' between the rotations $\bQ\spm$. This is related to the moment transferred by the base pair $\bc$ via an elastic constitutive relation for the web, 
\eqsp{
	&\bP=(\bQ\poss\bQ^{-T})^\frac{1}{2}=\bZ(\textbf{I}+\Phi^c)\bZ^T.}
Here, $\Phi^c=\frac{\bRp\Theta\poss\bR^{+T}-\bRn\Theta\negs\bR^{-T}}{2}$ is a skew symmetric tensor. The moment transferred by the base pair is directly proportional to the Gibbs vector of $\bP$. $\boeta=\tan \frac{\lambda}{2}\mathbf{\hat{k}}$ is a Gibbs rotation vector for a rotation matrix $\textbf{T}$ if $\textbf{T}\mathbf{\hat{k}}=\mathbf{\hat{k}}$ and $1+2\cos \lambda=$ tr$\textbf{T}$. In our case, the Gibbs vector of $\bP$ is $2\boeta=2\bZ\bar{\boeta}=\bZ\mathbf{\phic}$, where $\mathbf{\phi^c}$ is the axial vector of skew symmetric tensor ${\Phi^c}$. Note that in the reference state, $\boeta_0=0$ since $\bP_0=(\bQ_0\bQ_0^T)^{1/2}=\mathbf{I}$. The rigid rotation of the base pair $\bQ=\bP \bQ\negs$. Here
\eqsp{
	&\bQ=\bZ(1+\Phi)\bQ_0,}
and $\Phi=\frac{\bRp\Theta\poss\bR^{+T}+\bRn\Theta\negs\bR^{-T}}{2}$ is a skew symmetric matrix. The moment exterted by $+$ strand on the $-$ strand by means of the elastic web, $\bc$, is computed using $\bc=\bQ\mathbf{H}\bQ^T\boeta$ where $\mathbf{H}=$ diag$[H_1,~H_2,~H_3]$ are the elastic moduli. Now, we shift our focus to the extension and shear of the web. In the reference configuration the displacement between the two strands $\bw_0=\frac{\brp_0-\brn_0}{2}=a\be_r$. In the deformed configuration $\bw=\frac{\brp-\brn}{2}$. The force $\bfo$ exerted by $+$ strand on the $-$ strand is computed using $\bfo=\bQ \mathbf{L} (\bQ^T\bw-\bQ_0^T\bw_0)$ where $\mathbf{L}=$ diag$[L_1,~L_2,~L_3]$ are the elastic moduli. Complete expressions for $\bfo$ and $\bc$ are provided in the supplement.\\
We now consider the contributions from the stacking energy. The center line of the double helix $\be_3$ undergoes both twist $k_3$ and extension $\xi=\frac{\xi\poss+\xi\negs}{2}$. We associate a quadratic stacking energy $E_s=K_c k_3^2+K_e (\frac{\xi\poss+\xi\negs}{2}) ^2$ to penalize this change in the orientation of successive base pairs. 
Due to this energy, the base pairs exert a body force $\textbf{\textit{l}}$ and a body moment $\textbf{\textit{h}}$ on both $\pm$ strands which are given by
\eqsp{\textbf{\textit{l}}=K_e(\frac{\xi\poss+\xi\negs}{2}) \be_3, \quad\quad \textbf{\textit{h}}=K_c k_3\be_3.
	\label{Eq:stack_energy}
}  
Now we have all the ingredients for solving the governing differential equations of a birod.  
Substituting these quantities in the balance laws (eqn.\ref{Eq:gov_eqn}) gives us a set of 12 differential equations. The complete procedure for solving those equations is in the supplement, however we highlight crucial points here. It follows from the governing equations that $\beta\poss=\beta\negs(=\beta$ say), $n^c_3=n_1=n_2=0$. $\beta\poss=\beta\negs$ implies $\xi\poss=\xi\negs=(\xi$ say) thereby reducing 12 equations to 6 equations in 6 unknowns $r,\beta,k_3,n^c_{1,2},n_3$. We look for solutions of the form,
\eqsp{&r(x)=r_0e^{-\lambda x},\beta(x)=\beta_0e^{-\lambda x},~\xi(x)=\xi_0e^{-\lambda x},\\&n^c_1(x)=n^c_{10}e^{-\lambda x},~n^c_2(x)=n^c_{20}e^{-\lambda x},~n_3(x)=n_{30}e^{-\lambda x}.}	
We substitute this form into the governing equations (eqn. \ref{Eq:gov_eqn}) and obtain an eigenvalue problem in $\lambda$. In order to make further progress, we need the values of the elastic constants. We use $K_c=80$ pNnm$^2$, $K_e=600$ pN, $L_1=L_2=L_3=H_1=H_2=H_3=10$ pN. In the supplement, we show that these values yield the correct twist, stretch and twist-stretch coupling moduli for double stranded B-DNA. Solving for the eigenvalues $\lambda$ we get 
\eqsp{\lambda=\pm\zeta\pm i\omega,\quad\quad\zeta=0.32\text{nm}^{-1},}
and the solution for the strain parameters $y_1=r,y_2=k_3$ and $y_3=\beta$ is of the form:
\eqsp{
	y_i(x)=&A_1\mathbf{V}_1(i)e^{(-\zeta -i\omega)x}+A_2\mathbf{V}_2(i)e^{(-\zeta +i\omega)x}+B_1\mathbf{V}_3(i)e^{(\zeta -i\omega)x}+B_2\mathbf{V}_4(i)e^{(\zeta +i\omega)x}.
	\label{Eq:sol}
}
where $\mathbf{V}_j(i)$ is the $i^{th}$ component of the eigenvector corresponding to the eigenvalue in the exponent. Clearly, the decay length $\zeta$ is only a function of the elastic parameters of dsDNA, in agreement with the conclusion of Kim \textit{et al} \cite{kim_science}. Note that the strain parameters are exponentially decaying while oscillating with the period $\omega$ of the double helix. We impose the boundary conditions on $r$ and $\beta$ remembering that the displacements of the strands must be continuous. For a protein binding at $x=p$,
\eqsp{
	&\text{as }x\to\pm\infty\quad r(x),~\beta(x)\to 0,\\
	&\text{at }x=p\quad r(0)=r_0,~\beta(0)=\beta_0. 
}
We present the variation of $r$, $k_3$ and $\beta$ for a protein binding at $x=0$ for two different sets of boundary conditions in fig. \ref{fig:1prot1}. Notice the sinusoidal correlation between the local deformation of base-pairs which is in agreement with earlier work which used Monte Carlo simulations\cite{gu2015dna,xu2013modeling}.\\
We show the deformed shapes of the helices in the fig. \ref{fig:helixshape} for three cases: first when one protein binds at $x=0$, second when two proteins bind at $x=\pm1.5$ nm, and third when two proteins bind at $x=\pm3.5$ nm. The boundary condition for each protein is $r_0=0.2$ nm$,\beta_0=0$. We deliberately choose large values for $r_0$ and $\beta_0$ to distinguish the deformed shape from the reference shape. The large configuration changes near the site of protein binding ($x=0$) decay exponentially with distance. Note the strong overlap in the deformation fields when the distance between two proteins is $3$ nm compared to $7$ nm. This overlap results in an interaction energy between the two proteins which we subsequently quantify using eqn. (\ref{Eq:int_energy}). \\ 
We now compute the interaction energy $\Delta G$ for two proteins. The energy functional of the double helical rod is 
\begin{multline}
	E[r,\beta,k_3]=\frac{1}{2}EI(\kappa\poss)^2+\frac{1}{2}EI(\kappa\negs)^2+\sum_{i=1}^{3}\frac{1}{2}(L_i\Delta \bw_i^2+H_i\hat{\eta}_i^2)+K_c k_3^2+K_e\xi^2,\end{multline}
where $\hat{\mathbf{\eta}}=\bQ^T\mathbf{\eta}$ and $\Delta \bw=\bQ^T\bw-\bQ_0^T\bw_0$. Consider two proteins, $P_1$ and $P_2$ binding at $x=0$ and $x=p$. The interaction energy $\Delta G$ defined as,
\eqsp{\Delta G(p)= E_{12}^{\{0,p\}}-E_1^0-E_2^p,  \label{Eq:int_energy}}
where $E_{12}^{\{0,p\}}=E[r_{12},\beta_{12},(k_{3})_{12}]$ is the energy of two proteins binding onto DNA at $x=0$ and $x=p$, while $E_1^0=E[r_1,\beta_1,(k_{3})_1]$ and $E_2^p=E[r_2,\beta_2,(k_{3})_2]$ are the energies of a single protein binding at $x=0$ and $x=p$, respectively. We linearly superimpose the strain fields from one protein ($r_1$ and $r_2$, etc) to get the resultant strain field ($r_{12}$ etc) caused by two proteins simultaneously binding to DNA. 
\eqsp{r_{12}(x)=r_1(x)+r_2(x-p).}
We obtain $\beta_{12}$ and $(k_3)_{12}$ similarly. We compute the interaction energy $\Delta G(p)$ as a function of the distance between two proteins $p$ and plot it in fig.\ref{fig:DeltaGandrho} together with
experimental data from \cite{kim_science}. In excellent agreement with experiment~\cite{kim_science} and numerical simulations~\cite{gu2015dna}, $\Delta G$ decays exponentially while oscillating with the period of the double helix ($\sim 10 $ bp). We justify this variation of interaction energy for a simple case as follows. Consider a strain parameter $\delta(x)$ and the associated quadratic energy potential $\mathcal{E}[\delta (x) ]=\int_{-\infty}^{\infty}\frac{\delta^2(x)}{2}~dx$. Similar to our strain paramters in eqn. \ref{Eq:sol} let us assume $\delta(x)=Ae^{-b x}\cos(\mu x)$, then 
\begin{align}\mathcal{E}[\delta(x)]=\int_{-\infty}^{\infty}\frac{\delta^2(x)}{2}~dx=\frac{A^2(2b^2+\mu^2)}{4b(b^2+\mu^2)} \end{align}
$\mathcal{E}[\delta(x-p)]=\mathcal{E}[\delta(x)]$. Now the strain obtained by superposing two strain sources a distance $p$ apart are $\delta_2(x)=\delta(x)+\delta(x-p)$. The energy functional corresponding to $\delta_2(x)$ is
\begin{align*}\mathcal{E}[\delta_2(x)]=&\frac{A^2(2b^2+\mu^2)}{2b(b^2+\mu^2)}+A^2c_1e^{-bp}\sin(\mu p)+A^2c_2e^{-bp}\cos(\mu p)\\
	=&\mathcal{E}[\delta(x)]+\mathcal{E}[\delta(x-p)]+\Delta G\end{align*}
where $c_1=\frac{b^3}{2b\mu (b^2+\mu^2)}$ and $c_2=\frac{\mu(\mu^2+2b^2+pb^3+pb\mu^2)}{2b\mu (b^2+\mu^2)}$.
It is notable how the decaying sinusoidal behavior of the interaction energy $\Delta G$ follows naturally from the functional form of the strain parameters and their eventual superposition.\\
Next, we focus on the width of the groove since many proteins are known to change the width of the major/minor groove of DNA~\cite{kopka1985molecular,kim_science,hancock2013control}. We define the width of the groove, $g(x)$, as follows (we do not have a major/minor groove because $\alpha = \pi$):
\eqsp{g(x)=\br\negs.\be_3(x+\frac{\pi}{2\omega})-\br\poss.\be_3(x-\frac{\pi}{2\omega}).} 
Note that in the reference configuration the groove width $g_0=\frac{\pi}{\omega}=\frac{p}{2}$. We consider a protein binding at $x=0$ and compute the change in groove width $\rho(x) = g(x) - g_{0}$ for two sets of boundary conditions, $r_0=0,\beta_0=0.02$ and $r_0=0.02$ nm, $\beta_0=0$ (see fig.\ref{fig:DeltaGandrho}). The groove width $\rho$ decays exponentially with increasing distance from the binding site while oscillating with periodicity of the double helix. This characteristic decaying sinusoidal oscillation is documented in \cite{gu2015dna,xu2013modeling} and is also observed experimentally \cite{kim_science}. It has been proposed that this change in groove width could explain the sinusoidally decaying interaction energy (notice the similarity of the two panels in fig.\ref{fig:DeltaGandrho}) between two proteins bound to 
DNA because the binding energy of a protein binding to DNA could potentially depend on the groove width. However, we have arrived at the decaying sinusoidal variation of the interaction energy by computing the elastic energy stored in the birod without assuming any connection to the groove width.\\ 
To conclude, we have uncovered a mechanism for allostery in DNA using the theory of elastic birods. Our analysis ties together continuum theory \cite{maddocks_birod}, experiments \cite{kim_science} and numerical simulations \cite{drsata2014mechanical,gu2015dna}. Our computations indicate that the interaction energy (eqn. (\ref{Eq:int_energy})) for two proteins bound to DNA decays exponentially while oscillating with the period of the DNA double-helix. The decay length depends only on the elastic characteristics of the web while the oscillatory behavior is inherited from the underlying double-helical geometry. 
Our techniques based on a helical birod model could potentially be applied to other molecules which have a double helical geometry such as dsRNA, coiled-coil intermediate filaments, etc.  \\

We acknowledge support from NSF through grant number NSF CMMI 1662101 and NIH through grant number NIH R01-HL 135254.   

\bibliography{reference}{}
\bibliographystyle{plain}

\section{Supplement}

\subsection{Expressions for quantities in main text}
In this section, we give expressions for the various quantities used in the main text. In eqn. (\ref{Eq:Rframes}),
\eqsp{&\textbf{n}^\pm_0=\mp(\cosr\be_1+\sinr\be_2),\\
	&\textbf{b}^\pm_0=\mp\cos k (-\sinr\be_1+\cosr\be_2)+\sin k \be_3,\\
	&\textbf{t}^\pm_0=\pm\sin k (-\sinr\be_1+\cosr\be_2)+\cos k \be_3,\\
	& \Omega^\pm_0=\Omega_0=\omega \sin k.
}
We use the following formulas to compute $\textbf{n}\spm_0,\textbf{b}\spm_0,\textbf{t}\spm_0,\Omega\spm_0$ in the reference configuration and $\textbf{n}\spm,\textbf{b}\spm,\textbf{t}\spm,\Omega\spm$ in the deformed configuration: 
\eqsp{
	&\textbf{t}^\pm=\pardev{\br^\pm}{x}/|\pardev{\br^\pm}{x}, \quad\Omega^\pm=\textbf{t}^\pm_x.\textbf{t}^\pm_x,\quad \textbf{n}^\pm=\frac{1}{\Omega^\pm}\textbf{t}^\pm_x. 
}
We use $\beta=\frac{\beta\poss+\beta\negs}{2}$, $\beta^c=\frac{\beta\poss-\beta\negs}{2}$, $\bfv_1=\cosr\bd_1+\sinr\bd_2$ and $\bfv_2=-\sinr\bd_1+\cosr\bd_2$.
The expression for the moment transferred by the web, $\bc$, in eqn. (\ref{Eq:gov_eqn}) is
\eqsp{
	\bc&=\bQ\mathbf{H}\bQ^T\boeta\\
	&=H_1(-ak_3-\omega r_x-a\beta_x)\bfv_1+ H_2\frac{(-ak_{3x}-\omega r_{xx}-a\beta_{xx})}{\omega}\bfv_2\\
	&+ H_3 (\beta^c-\frac{a\cot k}{\omega}\beta^c_{xx})\be_3.
	\label{Eq:c}
}
The displacement between the two strands $\bw=\frac{\brp-\brn}{2}$ is
\eqsp{
	&\bw=(a+r)\bfv_1+a\frac{\beta\poss+\beta\negs}{2}\bfv_2+\frac{1}{2}\int_{-\infty}^{x}a(\xi\poss-\xi\negs)dx\be_3,\\
	&~~~=(a+r)\bfv_1+a\beta\bfv_2-a^2 \omega \beta^c\be_3.
}
We use inextensibility conditions from eqn. (\ref{Eqn:inext}) in the main text to evaluate the integral above:
\eqsp{
	\int_{-\infty}^{x}a(\xi\poss-\xi\negs)dx=-a^2\omega(\beta\poss-\beta\negs).
}
The expression for the force exerted by the $+$ strand on the $-$ strand, $\bfo$, in eqn. (\ref{Eq:gov_eqn}) is  
\eqsp{
	&\bfo=\bQ \mathbf{L} (\bQ^T\bw-\bQ_0^T\bw_0)\\
	&=L_1 r\bfv_1+aL_2\cot k \frac{ak_{3x}+2\omega r_x+a\beta_{xx}}{\omega}\bfv_2-L_3a^2\frac{\omega^2 \beta^c+\beta^c_{xx}}{\omega}\be_3
	\label{Eq:f}
}

\subsection{Elastic parameters for the web}
To estimate the elastic parameters of the web we do as follows.
We apply a stretching force $F$ and torque $T$ on one end of the double helix which changes the radius, twist and pitch. We assume these displacement variables are constant throughout the length of the deformed helix. Then we compute the elastic energy stored in the structure in terms of the strains and the unknown elastic moduli. Then we compute second derivatives of this elastic energy with respect to appropriate
strains to get expressions for the stretch modulus $S$, twist modulus $C$, twist stretch coupling modulus $g$ of the double helix in terms of the elastic parameters of the birod (web and strands). Then, 
we get estimates for the values of the elastic constants $L_{1,2,3}, H_{1,2,3}, K_e, K_c$ that will reproduce the known values of $S, C, g$ for DNA. To begin, the position vectors are: 
\begin{equation}
	\begin{split}
		&\brp=(a+r)(\cosr(1+\beta)\be_1+\sinr(1+\beta)\be_2)+x(1+e),\\
		&\brn=-(a+r)(\cosr(1+\beta)\be_1+\sinr(1+\beta)\be_2)+x(1+e),\\
	\end{split}
	\label{Eq:app_r0}
\end{equation} 
We use the following basis in the forthcoming calculations.
\eqsp{
	&\be_r=\cosr\be_1+\sinr\be_2,\\
	&\be_{\theta}=-\sinr\be_1+\cosr\be_2.
}
As usual we assume that $r,\beta,e\sim \oeps$ are small which allows us to linearize eqn. \ref{Eq:app_r0}. Also $r_x=\beta_x=\xi_x=0$.
\begin{equation}
	\begin{split}
		\brp&=(a+r)\be_r+a\omega \beta x\be_{\theta}+x(1+e)\be_3,\\
		\brp_x&=(a+r)\omega\be_{\theta}+a\omega\beta\be_{\theta}-a\omega^2\beta x\be_r+(1+(ex)_x)\be_3,\\
		&=-a\omega^2\beta x\be_r+\omega(a+r+a\beta)\be_{\theta}+(1+(ex)_x)\be_3.
	\end{split}
\end{equation}
The outer strands are inextensible, hence 
\begin{equation}
	\begin{split}
		&|\brp_x|=|\brp_{0x}|,\\
		&(ex)_x+\omega^2a(r+\beta)=0,\\
		& r=-\frac{e}{\omega ^2 a}-a\beta.
	\end{split}
	\label{Eq:appendix_inext}
\end{equation}
We follow the same steps as done in the main text to compute the director frame in the deformed configuration $\bRp$. The tangent vector in the deformed configuration $\tp$ is,
\begin{equation}
	\begin{split}
		\tp=&-\sin k \beta x\be_r+(\sin k+\omega r\cos k+\beta\sin k )\be_{\theta}+(\cos k-\omega \sin k(r+a\beta))\be_3,\\
		=&\tpz+\omega \beta x \sin k~\mathbf{n}_0\poss+(\omega r+\beta \tan k)\bpz,
	\end{split}
\end{equation}
We use the above expression to compute the change in curvature for the outer strands $\kappa \poss$. Differentiating the tangent vector with respect to $x$,
\begin{equation}
	\tp_x=-(\omega \sin k + 2\omega \beta \sin k+\omega^2 r \cos k)\be_r-\omega^2\sin k \beta x \be_{\theta}.\\
\end{equation}
We use the above expression to compute the curvature ($K$) of the outer strand in the deformed configuration. 
\begin{equation}
	\begin{split} 
		K=&\omega \sin k +2\omega \beta \sin k+\omega^2 r\cos k.\\
		\kappa\poss=&K-\omega\sin k=2\omega \beta \sin k+\omega ^2 r \cos k.\\
	\end{split}
\end{equation}
The expression for the normal in the deformed configuration $\np$ is given by, 
\begin{equation}
	\begin{split}
		\np=&-\er-\omega \beta x \be_{\theta},\\
		=&\np_0-\omega \beta x \sin k \tpz+\omega \beta x \cos k \bpz.
	\end{split}
\end{equation}
We are now in a position to calculate the deformed Frenet-Serret frame $\bRp$.
\eqsp{
	&\bRp=[\np\quad\bp\quad\tp]=\bRp_0(\mathbf{1}+\thetap)\\
		&\thetap=\begin{bmatrix}
			0 &-\theta^+_3&\theta^+_2\\\theta^+_3&0&-\theta^+_1\\-\theta^+_2&\theta^+_1&0
		\end{bmatrix},\\
		& \theta^+_1=\omega r+\beta \tan k, \quad  \theta^+_2=\omega \beta x \sin k,\quad \theta^+_3=\omega \beta x \cos k.
}
We replicate the procedure for the negative strand and get,
\begin{equation}
	\begin{split}
		&\bRn=[\nn\quad\bn\quad\tn]=\bRn_0(\mathbf{1}+\thetan),\\
		&\thetan=\thetap,\\
		&\kappa\negs=\kappa\poss.
	\end{split}
\end{equation}
The energy functional of the double helix is,
\begin{equation}
	\begin{split}
		E=&\int_{0}^{L}(EI(2\omega \beta \sin k+\omega^2 r\cos k)^2+\frac{1}{2}H_1\omega^2(r+a\beta)^2\\
		&+\frac{1}{2}L_1 r^2)\,dx-M\theta -F\Delta x,\\
		&\Delta x=eL,\quad\quad\theta=\beta L.\\
	\end{split}
\end{equation}
We eliminate $r$ using eqn. (\ref{Eq:appendix_inext}) and compute the elastic constants as follows. 
\begin{equation}
	\begin{split}
		&\frac{\partial E}{\partial \beta}=0,\quad\quad \frac{\partial E}{\partial e}=0.\\
		&S=\frac{\partial^2 E}{\partial e^2},\quad g=\frac{\partial^2 E}{\partial e\partial \beta},\quad C=\frac{\partial^2 E}{\partial  \beta^2}.
	\end{split}
\end{equation}
We use
\eqsp{
	&K_c=80 \text{ pNnm}^2,\quad K_e=600\text{ pN}, \quad H_1=H_2=H_3=10\text{ pN},\\
	&L_1=L_2=L_3=10 \text{ pN nm}^{-1},\quad EI=65 \text{ pN nm}^2.
	\label{eqn:elas_const} 
}
These elastic constants satisfy all the relevant experimental data. We obtain decay length $10$ bp \cite{kim_science}, S=$1243$ pN, C=$465$ pNnm$^2$ and g=-98 pNnm which are in the correct range \cite{singh2018allosteric}. We note that the choice of the elastic constants is not unique.
\subsection{Solution of governing equations}
Here we discuss the details regarding solution of the governing differential equations in the main text. Recall that the twelve equilibrium equations are: 
\begin{equation}
	\begin{split}
		&n_{1x}-\omega n_2=0,\\
		&n_{2x}+\omega n_1=0,\\
		&n_{3x}+K_e(\xi\poss_x+\xi\negs_x)/2=0,\\
		&n^c_{1x}-\omega n^c_2-f_1=0,\\
		&n^c_{2x}+\omega n^c_1-f_2=0,\\
		&n^c_{3x}-f_3=0,\\
		&EI\cos{k}(\kappa\poss-\kappa\negs)\omega-2n_2+2a\omega n^c_3=0,\\
		&EI\cos{k}(\kappa\negs_x-\kappa\poss_x)+2n_1-2af_3=0,\\
		&EI\sin k(\kappa\poss_x+\kappa\negs_x)+2af_2-2a\omega n^c_1+2K_c k_{3x}=0,\\
		&EI\cos{k}(\kappa\poss+\kappa\negs)\omega+2a\omega n_3-2n^c_2-2c_1=0,\\
		&-EI\cos{k}(\kappa\negs_x+\kappa\poss_x)+2n^c_1-2c_2=0,\\
		&EI\sin k(\kappa\poss_x-\kappa\negs_x)-2a\omega n_1-2c_3=0.
	\end{split}
	\label{eq:12diffeqns}
\end{equation}
We substitute the relevant quantities and get following set of equations where $\beta\upc=\frac{\beta\poss-\beta\negs}{2}$ and $\beta=\frac{\beta\poss+\beta\negs}{2}$.

\begin{subequations}
	\begin{equation}
		n_{1x}-\omega n_2=0,
		\label{goveqa}
	\end{equation}
	\begin{equation}
		n_{2x}+\omega n_1=0,
		\label{goveqb}
	\end{equation}
	\begin{equation}
		n_3+K_e (\xi\poss+\xi\negs)/2=0,
		\label{goveqc}
	\end{equation}
	\begin{equation}
		n^c_{1x}-\omega n^c_2-L_1 r=0,
		\label{goveqd}
	\end{equation}
	\begin{equation}
		n^c_{2x}+\omega n^c_1-aL_2\cot k\frac{a(k_{3x}+\beta_{xx}+2\omega r_x)}{\omega}=0,
		\label{goveqe}
	\end{equation}
	\begin{equation}
		n^c_{3x}+\frac{a^2L_3}{\omega}(\omega^2\beta\upc+\beta\upc_{xx})=0,
		\label{goveqf}
	\end{equation}
	\begin{equation}
		2a\omega n^c_3+2aEI\omega^2\cos^2 k\beta\upc_x=0,
		\label{goveqg}
	\end{equation}
	\eqsp{	&-2aEI\omega \cos^2 k\beta\upc_{xx}+\frac{1}{2}L_3(2a^3\omega \beta\upc+\frac{2a^3}{\omega}\beta\upc_{xx})=0,
		\label{goveqh}}
	\eqsp{
		&2EI\cos k \sin k (2a\omega (k_{3x}+\beta_{xx})+\omega ^2 r_x-r_{xxx})\\
		&+2a^2L_2\cot k\frac{a(k_{3x}+\beta_{xx})+2\omega r_x}{\omega}-2a\omega n^c_1+2K_ck_{3x}=0,
		\label{goveqi}}
	\eqsp{
		&2EI\omega\cos^2 k(2a\omega (k_{3}+\beta_{x})+\omega ^2 r-r_{xx})+2(-n^c_2+a\omega n_3)\\
		&+2H_1(a(k_3+\beta_x)+\omega r)=0,
		\label{goveqj}}
	\eqsp{&-2EI\cos^2 k(2a\omega (k_{3x}+\beta_{xx})+\omega ^2 r_x-r_{xxx}+2n^c_1+\\
		&\quad2H_2\frac{a(k_{3x}+\beta_{xx})+\omega r_{x}}{\omega})=0,
		\label{goveqk}}
	\eqsp{2aEI\omega\sin 2k \beta\upc_{xx}+\frac{H_3}{\omega}(-2\omega\beta\upc+2a\cot k \beta\upc_{xx}),
		\label{goveql}}
\end{subequations}

We see from eqn. (\ref{goveqa}) and (\ref{goveqb}) that $n_1=n_2=0$. Eqn. (\ref{goveqf}), (\ref{goveqg}), (\ref{goveqh}), (\ref{goveql}) give $\beta\upc=0$ and $n\upc_3=0$. This implies
\eqsp{
	\xi\poss-\xi\negs=a\omega(\beta\poss_x-\beta\negs_x)=0.
}
Hence, $\beta\poss=\beta\negs=\beta$ and $\xi\poss=\xi\negs=\xi$. The non-trivial set of equations are (\ref{goveqc}), (\ref{goveqd}), (\ref{goveqe}), (\ref{goveqi}), (\ref{goveqj}) and (\ref{goveqk}). We have six equations in six unknowns $r,\beta,k_3,n\upc_1,n\upc_2$ and $n_3$. As pointed out in the main text we substitute 
\eqsp{
	&r=r_0e^{-\lambda x},\beta=\beta_0e^{-\lambda x},k_3=k_{30}e^{-\lambda x},\\
	&n\upc_1=n\upc_{10}e^{-\lambda x},n\upc_2=n\upc_{20}e^{-\lambda x},n_3=n_{30}e^{-\lambda x}.
}
We use the elastic constants given in eqn. \ref{eqn:elas_const}. The resultant system can be expressed in the form of an eigenvalue problem in $\lambda$ with eigenvalues,
\eqsp{
	\lambda=\pm0.32\pm1.87i=\pm\zeta\pm i\omega \quad\text{(say).}
}  
Thus the solution vector $\bv(x)=[r(x)\quad k_3(x)\quad\beta(x)\quad n\upc_1(x)\quad n\upc_2(x)\quad n_3(x)]$ can be obtained as follows,
\eqsp{
	\bv(x)=&A_1(\bu_1-i\bv_1)e^{(-\zeta -i\omega)x}+A_2(\bu_1+i\bv_1)e^{(-\zeta +i\omega)x}+\\
	&B_1(\bu_2-i\bv_2)e^{(\zeta -i\omega)x}+B_2(\bu+i\bv_2)e^{(\zeta +i\omega)x}.
	\label{Eq:solvec}
}
Now substitute 
\eqsp{ A_1=\frac{A-i\tilde{A}}{2}\quad A_2=\frac{A+i\tilde{A}}{2},\\
	B_1=\frac{B-i\tilde{B}}{2}\quad B_2=\frac{B+i\tilde{B}}{2},\\	
}
into eqn.\ref{Eq:solvec} and get
\eqsp{
	\bv(x)=&e^{-\zeta x}\Big(A(\bu_1\cosr-\bv_1\sinr)-\tilde{A}(\bv_1\cosr+\bu_1\sinr)\Big)\\
	&e^{\zeta x}\Big(B(\bu_2\cosr-\bv_2\sinr)-\tilde{B}(\bv_2\cosr+\bu_2\sinr)\Big)
}
We use the following vectors for compact representation.
\eqsp{
	&\bw_1=\bu_1\cosr-\bv_1\sinr,\\
	&\by_1=\bv_1\cosr+\bu_1\sinr,\\
	&\bw_2=\bu_2\cosr-\bv_2\sinr,\\
	&\by_2=\bv_2\cosr+\bu_2\sinr.}
Now we can recover the expressions for our strain parameters $r(x),\beta(x)$ and $k_3(x)$ from the above equation.
\eqsp{
	&r(x)=e^{-\zeta x}\Big(A\bw_1(1)-\tilde{A}\by_1(1)\Big)+e^{\zeta x}\Big(B\bw_2(1)-\tilde{B}\by_2(1)\Big),\\
	&\beta(x)=e^{-\zeta x}\Big(A\bw_1(2)-\tilde{A}\by_1(2)\Big)+e^{\zeta x}\Big(B\bw_2(2)-\tilde{B}\by_2(2)\Big),\\
	&k_3(x)=e^{-\zeta x}\Big(A\bw_1(3)-\tilde{A}\by_1(3)\Big)+e^{\zeta x}\Big(B\bw_2(3)-\tilde{B}\by_2(3)\Big),\\
	&\xi(x)=-\omega^2 r(x)-a\omega(k_3(x)+\beta_x(x)).
}
We have four constants $A,\tilde{A},B$ and $\tilde{B}$. We need four boundary conditions to evaluate them. We impose the boundary conditions on $r(x)$ and $\beta$. The boundary conditions for a protein binding at $x=a$ are as follows,
\eqsp{
	&\text{as }x\to \infty\quad\quad r(x),\beta(x)\to 0,\\
	&\text{at }x= p\quad\quad r(x)=r_0,\quad \beta(x)=\beta_0.
}
When two proteins bind to DNA, one at $x=0$ and second at $x=p$ we superimpose the corresponding displacement fields from protein 1, $u^1$, and protein 2, $u^2$, to get resultant displacement field $u_2$. Here $u$ could be $r,\beta$ and $k_3$.
\eqsp{
	u_2(x)=u^1(x)+u^2(x-p).
}

\end{document}